# Multidimensional Visualization of Oracle Performance Using Barry007[†]

Tanel Põder[a] and Neil J. Gunther[b]

PoderC Pte. Ltd., Singapore[a]

Performance Dynamcis Company, Castro Valley, California, USA[b]

{tanel@poderc.com, njgunther@perfdynamics.com}

*Most generic performance tools display only system-level performance data using 2-dimensional plots or diagrams and this limits the informational detail that can be displayed. Moreover, a modern relational database system, like Oracle, can concurrently serve thousands of client processes with different workload characteristics, so that generic performance-data displays inevitably hide important information. Drawing on our previous work, this paper demonstrates the application of Barry007 multidimensional visualization to the analysis of Oracle end-user, session-level, performance data, showing both collective trends and individual performance anomalies.*

## 1 INTRODUCTION

Like most other graphical performance monitoring tools, the *performance visualization* (PerfViz) mindset in the Oracle database world is currently very two-dimensional. The typical performance dashboard (Fig. 1) comprises standard strip charts with time running along the $x$-axis and the aggregation of certain database metrics positioned on the $y$-axis. While this is often sufficient to get a rough idea of overall system performance, a dashboard view can be completely misleading [MM07]; especially when it comes to analyzing what individual Oracle processes are actually doing inside the database.

The purpose of this paper to demonstrate how a multitude of Oracle performance metrics can be compressed into 2- and 3-dimensional visualizations by applying various *barycentric coordinate* transformations; collectively referred to as *Barry007*. In previous work [Gun92, JG07, GJ07, Gun08], we have applied barycentric coordinates to visualizing multiprocessor utilization (*Barry3* coordinate system), web application response time data (*Barry3* coordinates) and network-segment utilization

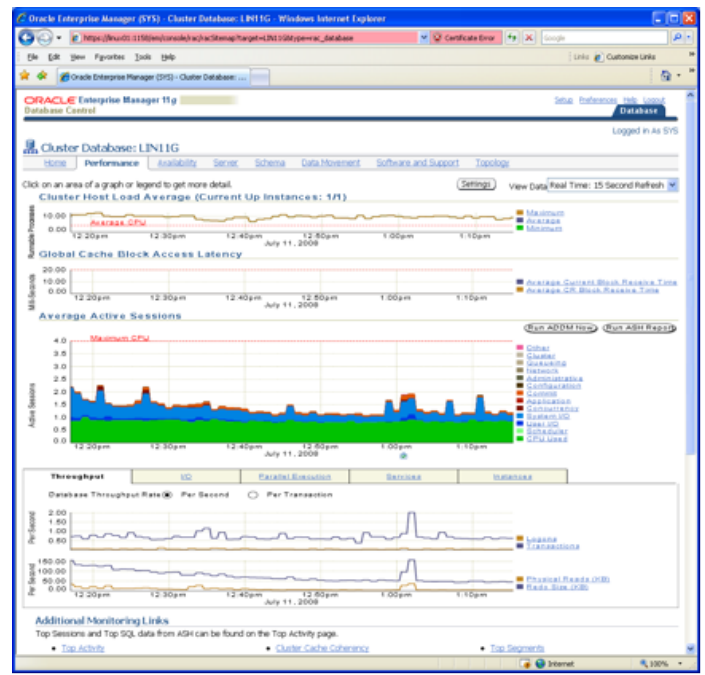

**Figure 1:** Oracle 11g Enterprise Manager

(*Barry4* coordinate system). The numeric value $N$ in the name *BarryN* indicates the number of performance metrics the can be visualized simultaneously. In the sub-

---

sequent sections, we report on the latest application of Barry007 coordinates to the *Oracle Wait Interface* metrics [Pod08], which are the primary performance indicators for Oracle database analysts.

## 2 CURRENT ORACLE INTERFACE

*Oracle Enterprise Manager*, shown in Fig. 1, is a performance monitoring tool commonly used by the Oracle database analyst (DBA). Its focus is on the database (DB) *response time* components and it facilitates easy data drill-down and navigation. However due to its 2-dimensional layout, it tends to obscure the visual identification of performance anomalies or performance trends of individual Oracle sessions or session groups. This limitation can lead to a situation where $1000$ application sessions with optimal performance can skew the system-wide statistics enough that the deleterious symptoms of $100$ suboptimal application sessions remain unnoticed.

Oracle maintains its performance data as relational tables in memory, known as v$ (pronounced "v-dollar") tables, which can be queried using SQL calls. Oracle has had session-level response-time instrumentation built in since early 1990s. It is called the *Oracle Wait Interface* (hereafter, OWI) and it presents a view of the wait timing information contained in the v$ tables.

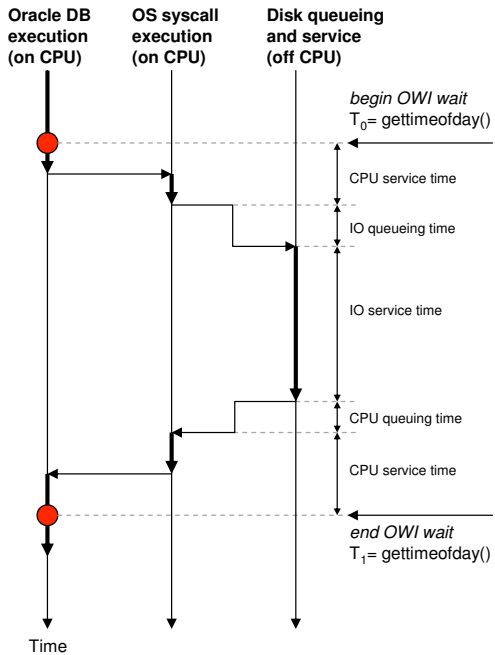

**Figure 2:** Definition of OWI waiting times

A word of caution may be appropriate here for those not familiar with Oracle OWI nomenclature. What is termed "wait time" in OWI is actually a *response time* [wJH03]. In Fig. 2, each OWI wait metric ($R_w$) is the sum of all actual wait times ($W_i$), in the sense of waiting for service, measured during $i$ intervals and the corresponding service times ($S_i$). More formally, we can write:

$$R_w = \sum_i W_i + \sum_i S_i \qquad (1)$$

There are some exceptions to this. For example, DB_CPU_PCT is purely a measure of processor service time (execution cycles) without any $W_i$ components. This is determined by what instrumentation and data sources are available to the OWI. See Appendix A for further discussion of this point.

The response time data is gathered at database-session level, allowing detailed performance analysis of individual database users. This presents many opportunities for performance diagnostics, such as deterministically diagnosing performance problems for individual DB users who are experiencing high latency, regardless of the fact that the database appears to be running efficiently. Collecting detailed performance information in the OWI helps to remove the mystery associated with performance problems by providing numerical evidence.

However, due an extra dimension in the collected data, viz., the need to view users and sessions together, we run into a limitation for effective data presentation. Two-dimensional plots, like those in Fig. 1, quickly become cluttered when there is more than just a handful of active individual sessions. If each Oracle metric is assigned to an independent coordinate axis, then the greater the number of metrics in the performance instrumentation dataset, the greater the number of coordinate axes required to view the result set.

One answer to this problem is to provide the visualization tool with an interactive capability, that allows the analyst to navigate around in the data in real time, drill down, and slice-and-dice the data as desired. Essentially, one would just observe the data from multiple viewing angles. This is equivalent to creating new dimensions into the visualization by representing it multiple times over time. Even with that capability, such tools would still only be reactive, in the sense that the performance analyst or DBA would need to be aware of the problem first, before applying the tool from various viewing angles to "look into" and thereby diagnose the issue. A classic *Catch-22* situation. Nonetheless, this approach is very similar to the first implemented in Tukey's PRIM-9 [Tuk88] tool, later modernized as MacSpin [AWD88] and now embedded in tools like Mathematica.

With these limitations as motivation, we now turn to the dimensional compression offered by barycentric coordinates. In the next section, we briefly review the concepts underlying the barycentric coordinate system. The interested reader can find more background in Ref. [JG07].

## 3 BARYCENTRIC COORDINATES

We introduce the concept of barycentric coordinates by showing that the locus of a *point* in the plane (i.e., two dimensions), defined by barycentric coordinates, is bounded by sides of an equilateral triangle, assumed to have height $h = 1$ for simplicity.

Referring to Fig. 3, the location of any point *inside* the triangle can be determined by the *lengths* of the three arrows perpendicular to each side of the triangle. These arrows are the barycentric coordinates. Identifying each arrow-length by $p_1$, $p_2$ and $p_3$, the *centroid* in Fig. 3(a) corresponds to the height divided into three equal lengths: $p_1 = p_2 = p_3 = 1/3$. The centroid is the center-of-mass or balance point if equal weights were to be placed at each vertex.

If instead, the weights were *unequal*, then the balance point would need to be shifted towards the heavier weights. As shown in Fig. 3(b), the sum of arrow-lengths still equals the height of the triangle:

$$p_1 + p_2 + p_3 = h = 1 \qquad (2)$$

Equation (2) is called a *sum rule* and it applies to *any* point interior of the triangle. It is an *invariant* because the three arrows partition the area of the triangle into three subareas that must, by definition, sum to the total triangle area.

The barycentric coordinates for $N$ performance metrics or $N$ degrees of freedom is a $d = N - 1$ *simplex*. For example, $N = 3$ metrics can be represented in a 2-dimensional equilateral triangle because such a triangle is a $d = (3-1) = 2$-simplex. $N = 4$ metrics corresponds to a 3-simplex or *tetrahedron* (See Fig. 8). The virtue of the simplex-based coordinate system is that is contains the extra degree of freedom for free! This is where some of the compression, mentioned in Sect. 2, comes from.

In general, a $d$-simplex has $N = d + 1$ faces. The corresponding $N$ barycentric coordinates are defined as the lines perpendicular to each face. Each vertex $Vi$ in Fig. 3 has Cartesian coordinates $(V_{ix}, V_{iy})$, where $i = 1, 2, 3$. If we choose the Cartesian coordinates $(1/\sqrt{3}, 1)$, $(0, 0)$ and $(2/\sqrt{3}, 0)$ for $V_1$, $V_2$, and $V_3$, respectively, then the

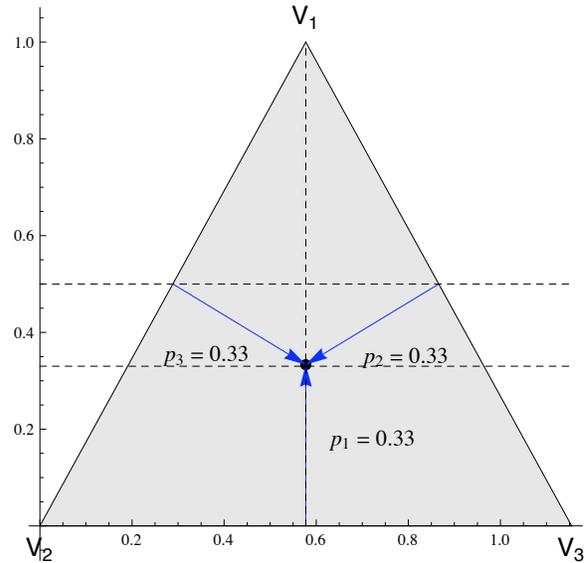

(a) The centroid or *center-of-mass* is the balance point of the triangle that would pertain if equal weights were placed at each vertex. It sits at a distance $h/3$ from each side. Hence, the barycentric coordinates are $p_1 = p_2 = p_3 = 0.3333$ when $h = 1$

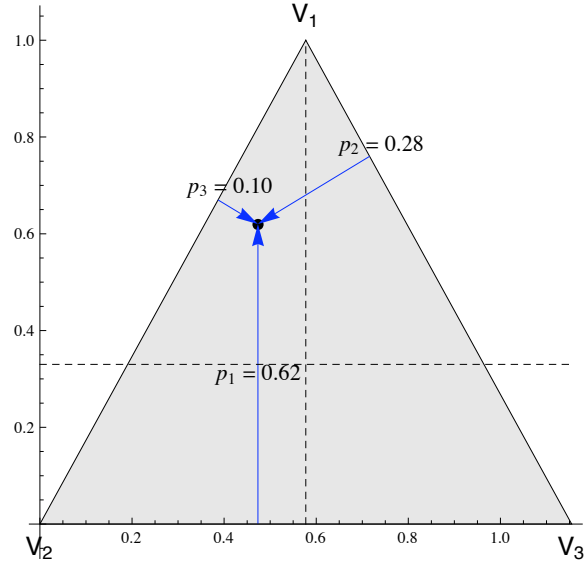

(b) If the vertex weights are different, the point $p$ must be relocated to compensate for this imbalance. Here the vertices $V_1$ and $V_2$ are "heavier" than $V_3$, so the barycentric coordinates become $p_1 = 0.62$, $p_2 = 0.28$, $p_3 = 0.10$ with $h = 1$. The dashed lines intersect at the centroid

**Figure 3:** Barry3 coordinates for a 2-simplex.

Cartesian coordinates of the point $(p_x, p_y)$ are given by:

$$\begin{aligned}(p_x, p_y) = (&p_1 V_{1x} + p_2 V_{2x} + p_3 V_{3x},\\ &p_1 V_{1y} + p_2 V_{2y} + p_3 V_{3y}).\end{aligned} \qquad (3)$$

Consider the centroid in Fig. 3(a). Its barycentric coor-dinates are $p_1 = p_2 = p_3 = 1/3$, so its corresponding Cartesian coordinates are:

$$(p_1V_{1x} + p_2V_{2x} + p_3V_{3x},\ p_1V_{1y} + p_2V_{2y} + p_3V_{3y})$$
$$= \left(\frac{1}{3}\frac{1}{\sqrt{3}} + 0 + \frac{1}{3}\frac{2}{\sqrt{3}},\ \frac{1}{3} + 0 + 0\right)$$
$$(p_x, p_y) = \left(\frac{1}{\sqrt{3}},\ \frac{1}{3}\right).$$

This result can be checked in Fig. 3(a). The vertex $V_1$ is located at $1/\sqrt{3} = 0.5774$ on the $x$-axis, so the Cartesian coordinates of the centroid are $(p_x, p_y) = (0.58, 0.33)$. Similarly, in Fig. 3(b), $p_1 = 0.62$, $p_2 = 0.10$ and $p_3 = 0.28$, so the three arrows meet at $(p_x, p_y) = (0.47, 0.62)$.

## 4 OWI METRICS TEST CASE

We now turn to the representation of OWI wait-class metrics in Barry3 coordinates. As proof-of-concept, we consider first aggregating several OWI metrics into just the three metrics needed to define the Barry3 axes: $p_1$, $p_2$ and $p_3$. The following eight OWI Oracle wait classes:

1. USERIO_PCT
2. SYSTEMIO_PCT
3. APPLICATION_PCT
4. COMMIT_PCT
5. CONCURRENCY_PCT
6. CONFIGURATION_PCT
7. NETWORK_PCT
8. OTHER_PCT

along with DB_CPU_PCT, were aggregated into 3 composite classes:

1. WAIT_PCT = APPLICATION_PCT + COMMIT_PCT + CONCURRENCY_PCT + CONFIGURATION_PCT + NETWORK_PCT + OTHER_PCT
2. IO_PCT = USERIO_PCT + SYSTEMIO_PCT
3. DB_CPU_PCT

shown as a stacked chart in Fig. 4(a). All of these metrics are drawn from the `v$system_wait_class` table and have been normalized to percentages (PCT).

Fig. 4(a) verifies graphically that the three composite OWI-test metrics, DB_CPU_PCT, IO_PCT and WAIT_PCT, do obey the Barry3 sum rule (2).

A performance analyst, however, is more likely to want to see these data clearly separated into their respective components, rather being stacked as in Fig. 4(a).

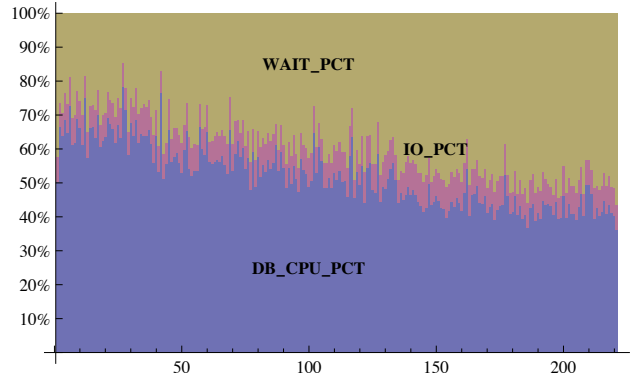

(a) Stacked-chart of the three composite OWI-test metrics plotted as a function of snapshots identified by a SNAP_ID

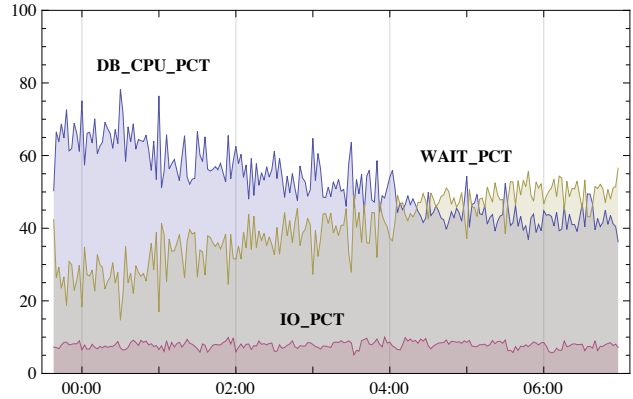

(b) OWI test metrics in Fig. 4(a) displayed as a time series

**Figure 4:** Composite OWI metrics for test case

Fig. 4(b) shows the same three OWI combined metrics plotted as a time series. We observe immediately that the IO_PCT metric remains relatively constant at around $10\%$ over the entire measurement period ($T \simeq 7$ hours). The DB_CPU_PCT metric is falling slowly, while WAIT_PCT increases proportionately. The latter metrics cross one another at about $4.5$ hours into the measurement window.

## 5 BARRY3 OWI REPRESENTATION

We are now in a position to transform the OWI test metrics of Fig. 4(a) into a Barry3 representation. The Barry3 axes associated with Fig. 3 are:

1. DB_CPU_PCT: "north-pointing" red arrow
2. WAIT_PCT: green arrow pointing "south-east"
3. IO_PCT: blue arrow pointing "south-west"

The corresponding Barry3 representation is shown in Fig. 5. This particular choice of axes is entirely arbitrary and any actual choice should be at the disposal of the user. Each SNAP_ID sample in Fig. 4(a) corresponds to the location of the dot in Barry3. Fig. 5(a) shows such a point with its barycentric coordinates $(p_1, p_2, p_3)$ corresponding to SNAP_ID $= 8$.

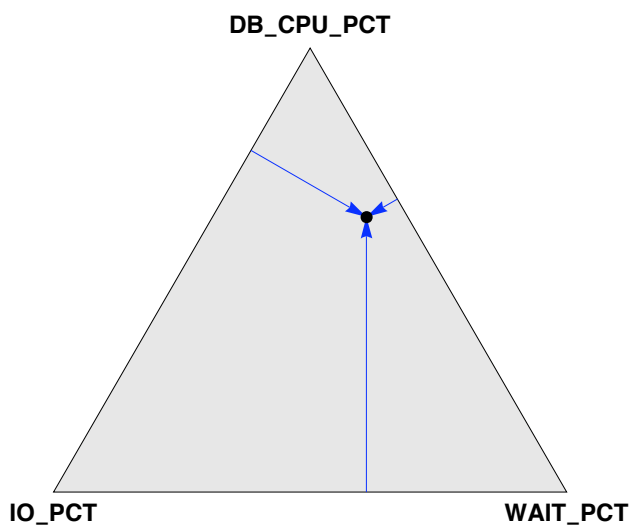

(a) SNAP_ID sample number 8 of the composite OWI-metrics in Fig. 4(a) represented by a dot inside the Barry3 triangle

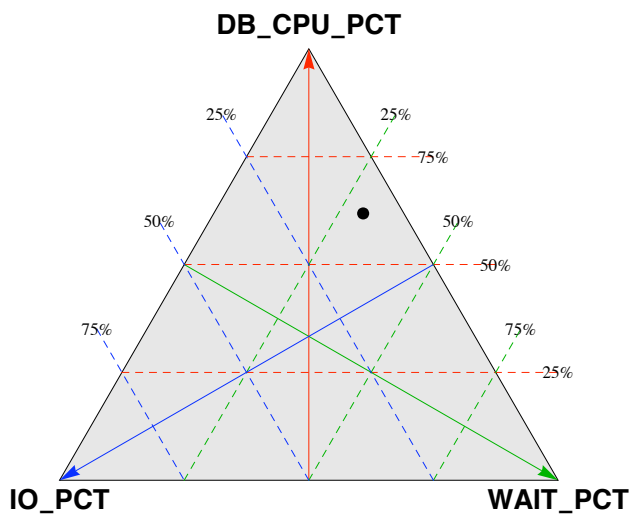

(b) Trilinear coordinates are a visual aid for assessing the contribution of each OWI-metric to the dot position inside Barry3

**Figure 5:** Single snapshot of composite OWI-metrics from Fig. 4(a) displayed as a dot in a Barry3 representation. A succession of such snapshots produces animation

We have chosen to make DB_CPU_PCT the *apex* of the Barry3 triangle because it is presumably more desirable to have Oracle processes executing than it is to have them waiting on (non-computational) resources. But this is matter of personal preference when it comes to visualization and beyond the prototype display shown here, it should be a user-definable option in any released PerfViz tool.

In Fig. 5(b) the barycentric coordinates in Fig. 5(a) have been replaced by *trilinear coordinates* [Har99]. Such a static coordinate system might be preferred for qualitatively assessing the contribution of each OWI-metric to the dot position.

A succession of OWI snapshots from Fig. 5 produces *animation* in Barry3. See http://www.perfdynamics.com/Test/owiCompB3.gif). The dot moves inside the triangle. Each time-step is displayed as an *odometer* value in the lower part of the triangle. As we soon see in an animation, the dot stays close to the edge of the Barry3 triangle. This corresponds to the constant $10\%$ IO_PCT time seen in Fig. 4(b). Gradually, the dot drifts down, along the edge, as the percentage of WAIT_PCT time increases at the expense of DB_CPU_PCT time.

# 6 MULTIPLE OWI SESSIONS

Next, we extend the proof-of-concept described in Sects. 4 and 5 to render *multiple* Oracle sessions in Barry3. For the purpose of Oracle data collection and cross-checking, we used the following tools:

**Sesspack:** Sesspack is a session-level data collection tool for Oracle. It samples relevant v$ tables periodically and stores the output to persistent tables for future analysis. The benefit of Sesspack relative to the usual Oracle data collectors lies in the level of detail captured and its flexibility to allow the analyst to specify both the sessions to monitor and the metrics to measure.

**PerfSheet:** PerfSheet is an Excel-based PerfViz tool which facilitates a convenient visualization of the result set of any SQL query, including those made against Sesspack data. The relevant data is automatically fetched to PerfSheet, after which a user is free to view that data in Excel pivot charts as desired.

See Ref. [Pod08] for more details. In the current prototypes presented here, the primary source of OWI data for Barry007 is Sesspack. The Barry3 axes selected from the available OWI metrics are:

1. CPU_USAGE: Oracle session is executing on a CPU.
2. IDLE: Session is idle waiting for next client request.
3. DB_WAIT: Other waits for I/O, DB locks, etc.

Note, that although these are not reported as percentages (as in Sect. 5), all three times must sum to the sample interval of 1000 ms, or 1 second, for each session such that:

$$\text{CPU\_USAGE} + \text{IDLE} + \text{DB\_WAIT} = 1000 \text{ ms} \quad (4)$$

in agreement with (2). An example data set with 60 simultaneous OWI sessions, displayed in Barry3 coordinates, is shown in Fig. 6.

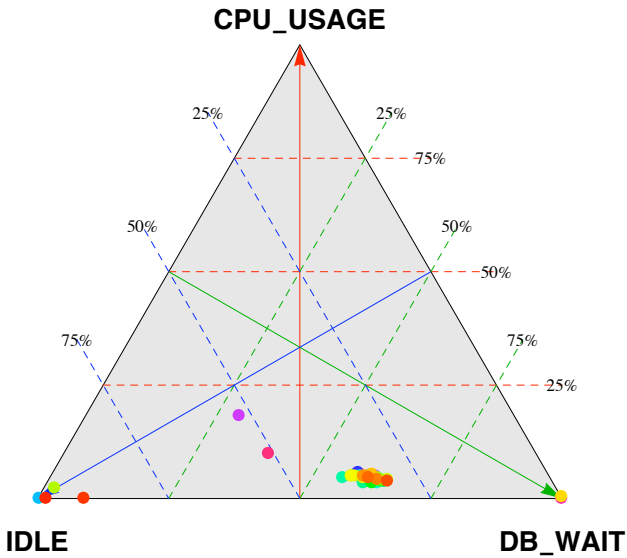

**Figure 6:** A snapshot of 60 OWI sessions displayed simultaneously as colored dots in Barry3. Note the slightly different choice of OWI metrics from those in Fig. 5

There is an important difference between the datasets used here and those in Sect. 4. The test-case OWI data do not contain an IDLE component. Therefore, we can only visualize the components of time that the DB spends servicing a user call viz., the "DB Time," in Oracle parlance. It is not possible to directly map this back to wall-clock time, because we do not know how much idle time was incurred.

In the session-level data being discussed here, the IDLE time is measured, so we can directly map the measured times to wall-clock time and any "missing" or multiply-accounted time indicates that some measurement error was involved. We shall return to the issue of measurement errors in Sect. 8.

In Fig. 6 (one frame of an animated time series) we see immediately that 4 sessions are completely idle (*lower-left vertex*), 2 sessions are 50% idle (*dashed-diagonal blue line second from left*), a cluster of sessions that are about 10% CPU busy (*group of colored dots at bottom center-right*), and 2 sessions are waiting on the DB for 100% of the time (*lower-right vertex*).

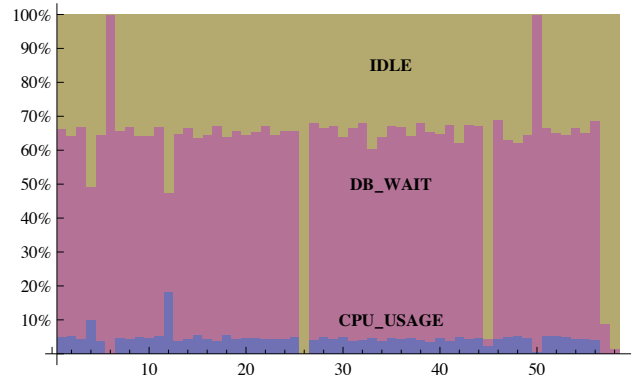

**Figure 7:** Stacked view of Fig. 6

Note that we just counted 2 points at the DB_WAIT_PCT vertex in Fig. 6, whereas only a *single* colored dot can be seen. This visual anomaly raises an important point; one that would have to be addressed in any real Barry3 visualizations. In this prototype display, one dot overlaps the other. We can verify this effect with the stacked view in Fig. 7. The DB_WAIT metric (shown in mauve) has two spikes that reach 100%. In an animated Barry3 display, overlapping points are less of an issue because the points are in constant motion. In a static Barry3 view, although overlapping is more significant, it could be corrected by adding a small offset to the positions of any overlapping dots.

## 7 BARRY4 OWI REPRESENTATION

We can further resolve the DBwait time in waiting for DB locks and latches and waiting on other resources e.g., physical I/O. The 4 OWI metrics then become:

1. CPU_USAGE: Oracle session is active and executing on a CPU.
2. DB_CONTENTION: Session is active but waiting on database locks or latches.
3. DB_WAIT: Session is active, but waiting on other waits like disk and network I/O.
4. IDLE: Session is inactive and waiting for next client request.

so we need to consider a Barry4 representation. Here, *active* means that the Oracle session is serving a user request.

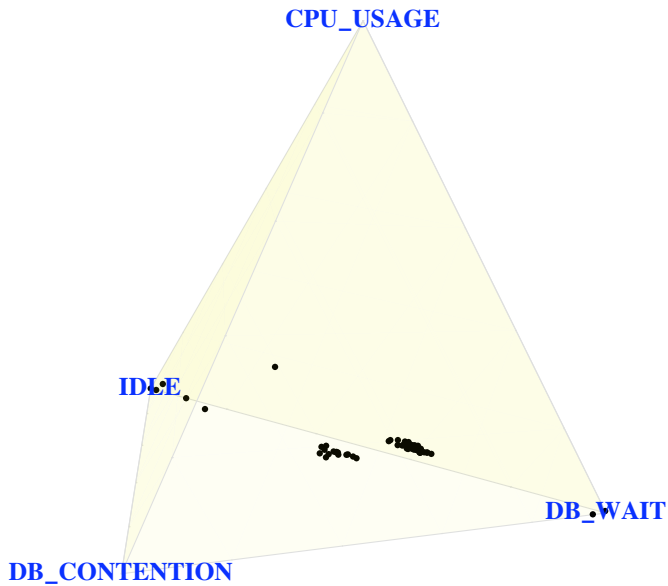

**Figure 8:** Snapshot of 60 OWI sessions in Barry4. Two clusters of points stand out at this viewing angle

Although these OWI metrics are not reported as percentages, they must sum to the sample interval of $1000$ ms or $1$ second in order to meet the sum rule condition (2).

$$\text{CPU\_USAGE} + \text{DB\_CONTENTION} + \text{DB\_WAIT} + \text{IDLE} = 1 \text{ s} \quad (5)$$

Barry4 is a $3$-simplex or tetrahedron shown in Fig. 8. We see that $2$ sessions are essentially inactive (near $100\%$ IDLE). There are two clusters of points that are very noticeable. One of the clusters is experiencing a moderate degree of DB_CONTENTION, while the other cluster is experiencing a moderately high degree of DB_WAIT time. In a presentation we would be able to demonstrate how the tetrahedron in Fig. 8 can be swiveled with the mouse to facilitate views from different perspectives.

## 8  OWI INSTRUMENTATION ISSUES

In order to build a multidimensional performance visualization solution which can make sense out of the large amount of instrumentation data, we need to focus on the end user response time. Simply monitoring the total response time, however, is not sufficient for diagnosing causes of performance problems, thus we need response time breakdown to individual events like CPU time, IO waits, DB lock waits, etc.

We also need full response time accounting, such that no component of the user response-time is missing from the total. We also need DB session level instrumentation with the ability to take snapshots of individual sessions and Sesspack [Pod08] provides that data for the Barry007 representation.

Since Barry007 imposes the constraint of a sum rule (2), it also exposed previously unseen limitations in Oracle's performance instrumentation viz., the OWI metrics in (4) and (5) do not always consistently add to 1000 ms. There are occasional bugs with OWI response time components unaccounted for. In addition, CPU preemption and run-queue waiting time in a multiprocessor or multicore are not accounted by default, across all platforms. So here is an unexpected diagnostic benefit of PerfViz. Without the aid of good visualization tools such defects can easily go unnoticed and continue to act as an unknown error source in any performance analysis and capacity planning.

The latest version of OWI reports 959 different wait events, which Oracle has grouped into a smaller number of wait classes:

```
SQL> select wait_class, count(*)
  2  from v$event_name
  3  group by wait_class;

WAIT_CLASS          COUNT(*)
------------------- ----------
Concurrency               26
System I/O                23
User I/O                  22
Administrative            51
Other                    630
Configuration             21
Scheduler                  3
Cluster                   47
Application               15
Queueing                   4
Idle                      80
Network                   35
Commit                     2
```

We would simply need to further collapse these wait classes in order to provide the appropriate input for Barry007 analysis.

## 9  CONCLUSION

In this paper we have presented another application of performance visualization based on the dimensional com-pression offered by barycentric coordinates. The sum-rule (2) imposes a very severe constraint on the performance metrics that can be displayed in the Barry007 style and this makes it challenging to find appropriate applications. As we have shown here, we can now add Oracle Wait Interface (OWI) to a growing list of metrics that are amenable to both Barry3 and Barry4 display.

Inaccuracies in collected performance data is an often overlooked subject because there is a widespread tendency to believe measured numbers and unless one has the right tools and techniques, the process of determining the level of measurement error can be a very difficult one. Another benefit of our approach was the ability to visually identify missing times in OWI. Without the aid of good PerfViz tools such defects can remain unnoticed and continue to introduce errors into any database performance analysis. With tools like Barry007 that enable data discovery rather than just data reporting, measurement error is more easily revealed and quantified.

By comparison with standard 2-dimensional tools that simply display time on the x-axis, Barry007 can apply animation to represent the time-development of each performance metric, thus freeing up the x-axis to display a different metric. Using current performance tools, the DBA is often forced to address each of the following performance issues separately:

1. Display the current performance state of a database with session-level detail
2. Early detection of performance anomalies
3. Detecting application-level performance trends

Barry007 could have a positive impact in Oracle performance management since it provides a solution for all three of these issues.

More generally, in the arena of PerfViz, a stalemate exists between performance-tool vendors who are disinclined to invest in engineering development when they do not see any demand, and performance analysts who are not aware of what they are missing in terms of better PerfViz for solving performance management problems.

All the illustrations and demonstrations in this paper were created with Mathematica 6.0 and are therefore only prototypes. Naturally, prototypes suffer from the kinds of limitations discussed in Sects. 6 and 7. Nonetheless, our hope is that these prototypes, as well as others [JG07], will provide further incentive to improve the kind of exploratory PerfViz tools available to both Oracle DBAs in particular and performance analysts in general.

## 10 ACKNOWLEDGMENTS

We thank Peter Stalder for providing the composite Oracle OWI data used in Sections 4 and 5.

## A APPENDIX

Here, we discuss in more detail Oracle's approach to decomposing application (end user) response time into its fine-grain components. Some examples where this is done, include: Oracle server process waiting for IO to be completed, Oracle waiting for a lock to be released and time spent executing on a CPU.

### A.1 OWI Wait Time

The wait times are measured inside Oracle. Whenever Oracle is about to issue a system call which could block and cause the process go off CPU, Oracle gets a system timestamp using gettimeofday() just before issuing the system call. Another timestamp is taken when a system call has completed and Oracle process has got back onto CPU. This has the following implications:

- Some of the Oracle "wait" time includes time spent executing on CPU as issuing the system call requires some CPU cycles. Also, the CPU busy time spent by OS serving the system call (in kernel mode) is also attributed to the Oracle process. Also, sometimes (like `read()` system calls from OS filesystem cache) do not cause the process go off CPU at all as the read can be satisfied from server main memory. In such cases Oracle reports short waits, while all this time was actually spent on CPU.
- Another implication is that if a process has been scheduled off CPU due blocking system call, then it needs to get back onto CPU to take the wait end timestamp. If there is some scheduling latency, it will be implicitly attributed to the wait event as well, since Oracle doesn't know about scheduling latency on most platforms.

### A.2 OWI CPU Time

On a Unix system, an Oracle server process executes `getrusage()` system calls at the end of every database call (or every 5 seconds for long DB calls) for getting CPU usage numbers for that process from OS. CPU scheduling latency is not included in Oracle's CPU time statistics as most OS-es don't account this. As CPU and Wait times are gotten using different methods, from different sources, this leaves some room for double counting errors.

## A.3 Barycentric Sum Rule

In the total (wall-clock) elapsed time, we have three components:

1. Oracle measured waits: the difference between Oracle's *wait begin* and *wait end* timestamps.

2. OS measured process CPU time usage. Oracle gets this data from OS. It can be interpreted as the *minimum* elapsed time Oracle spent in CPU service.

3. Unaccounted time ($T_u$). This includes all other waits:
$$T_u = R - W_O - B$$
where $R$ is the end-to-end user response time, $W_O$ is the wait-time as measured by the Oracle RDBMS, and $B$ is CPU busy-time as measuredby the OS. $T_u$ includes scheduling latency, measurement errors, instrumentation bugs.

**TRADEMARKS**